# In vivo quantification of 3D displacement in sacral soft tissues under compression: Relevance of 2D US-based measurements for pressure ulcer risk assessment

**Authors:**

**Ekaterina Mukhina***
Univ. Grenoble Alpes, CNRS, TIMC, 38000 Grenoble, France.
ekaterina.mukhina@univ-grenoble-alpes.fr

**Alessio Trebbi**
Univ. Grenoble Alpes, CNRS, TIMC, 38000 Grenoble, France.
Alessio.Trebbi@univ-grenoble-alpes.fr

**Pierre-Yves Rohan**
Institut de Biomécanique Humaine Georges Charpak, Arts et Métiers ParisTech, 151 bd de l'Hôpital, 75013. Paris, France
Pierre-Yves.ROHAN@ensam.eu

**Nathanaël Connesson**
Univ. Grenoble Alpes, CNRS, TIMC, 38000 Grenoble, France.
nathanael.connesson@univ-grenoble-alpes.fr

**Yohan Payan**
Univ. Grenoble Alpes, CNRS, TIMC, 38000 Grenoble, France
Yohan.Payan@univ-grenoble-alpes.fr

## Abstract

*Objective:* 2D Ultrasound (US) imaging has been recently investigated as a more accessible alternative to 3D Magnetic Resonance Imaging (MRI) for the estimation of soft issue motion under external mechanical loading. In the context of pressure ulcer prevention, the aim of this pilot MRI study was to design an experiment to characterize the sacral soft tissue motion under a controlled mechanical loading. Such an experiment targeted the estimation of the discrepancy between tissue motion assessed using a 2D imaging modality (echography) versus tissue motion assessed using a (reference) 3D imaging modality (MRI).

*Methods:* One healthy male volunteer participated in the study. An MRI-compatible custom-made setup was designed and used to load the top region of the sacrum with a 3D-printed copy of the US transducer. Five MR images were collected, one in the unloaded and four in the different loaded configurations (400-1200 [g]). Then, a 3D displacement field for each loading configuration was extracted based on the results of digital volume correlation. Tissue motion was separated into the X, Y, Z directions of the MRI coordinate system and the ratios between the out-of-plane and in-plane components were assessed for each voxel of the selected region of interest.

*Results:* Ratios between the out-of-plane and in-plane displacement components were higher than 0.6 for more than half of the voxels in the region of interest for all load cases and higher than 1 for at least quarter of the voxels when loads of 400-800 [g] were used.

*Conclusion:* The out-of-ultrasound-plane tissue displacement was not negligible, therefore 2D US imaging should be used with caution for the evaluation of the tissue motion in the sacrum region. The 3D US modality should be further investigated for this application.

## Introduction

Internal soft tissue deformation has been shown to be one of the main factors responsible for the onset of Pressure Ulcers (PU) and to be representative of the risk of the PU development. Indeed, previous work in animal models have established that compression-induced damage and internal tissue strains are correlated (Gawlitta et al. 2007; S. S. Loerakker et al. 2010; S. Loerakker et al. 2011; Nelissen et al. 2019; Stekelenburg et al. 2006; Traa et al. 2018; 2019; van Nierop et al. 2010). Based on the results obtained on N=11 female Brown-Norway rats, Ceelen et al. established that tissue damage could be measured using T2-weighted MRI when the maximum shear strain in the tissues and the compressive strain were in excess of 75% and 45 % respectively (Ceelen et al. 2008). The experimental quantification of soft tissue displacements and associated strain fields when tissues are compressed is therefore an important question. Monitoring strain fields could potentially be used to provide a quantitative metric to assist in the clinical evaluation of injury risk. However, the *in vivo* (and more challenging, *in situ*) monitoring of mechanical strains represents a significant challenge for the community.

B-mode ultrasound (US) imaging has been shown to be promising for the quantification of soft tissue motion in combination with Digital Image Correlation (DIC) (Gennisson et al. 2013). It has been used to quantify soft tissue displacements *in vitro*, in tissue-mimicking phantom (Zhu et al. 2015), *ex vivo*, in porcine flexor tendon (Chernak Slane and Thelen 2014) and, *in vivo*, in the human Achilles tendon (Chimenti et al. 2016) and in the quadriceps muscle (Affagard, Feissel, and Bensamoun 2015). From the perspective of PU prevention, a recent study by Doridam et al. (Doridam et al. 2018) investigated the feasibility of using B-mode ultrasound imaging combined with DIC for the quantification of subdermal soft tissue strains in the buttock region in two perpendicular planes (sagittal and frontal) during sitting. Results showed that, in both planes, the muscle tissue motion in the second principal direction (perpendicular to the pelvis motion) was important suggesting there was a non-negligible out-of-ultrasound-plane motion of the material particles. As a result, tracking muscle features using image registration techniques in each plane would introduce biases. This is a major limitation for the use of 2D US imaging for the *in vivo* assessment of soft tissue motion under mechanical loading.

Other attempts have been made using Magnetic Resonance Imaging (MRI) associated with Digital Image Correlation (DIC) to quantify tissue motion. MRI associated with cross correlation techniques is actually considered the gold standard for the assessment of 3D tissue motion under mechanical loading (Gilchrist et al. 2004; Bay 2008). From a PU prevention perspective, Solis et al. assessed the internal displacements and the associated strain fields *in vivo* in healthy and spinal cord injury (SCI) pigs using tagged MRI observing an increase in the values of shear and tensile strains with an increase in the distance from the centre of ischial tuberosity ventrally (Solis et al. 2012). In humans, Sonenblum et al. used 3D seated MRI to evaluate displacements in buttock tissues during sitting in able-bodied and SCI subjects. Results showed that 5 out of 7 tested subjects did not have notable muscle tissue under the ischial tuberosity while sitting suggesting the importance of using multi-planar imaging to assess subject-specific anatomy (Sonenblum et al. 2015). Likewise, Trebbi et al. combined MRI of the foot in both deformed and undeformed configurations with Digital Volume Correlation (DVC), to experimentally assess internal tissue displacements in the heel pad region (Trebbi et al. 2021) under external shearing load. However, although MRI is a potential tool for the quantitative evaluation of soft tissue displacements, it has important drawbacks such as high costs, a confined environment and the requirement for the patient to not move for long periods in positions that can be uncomfortable.

To summarize, ultrasound-based investigations represent a promising alternative to MRI-based assessment of tissue motion in clinical environment because they improve on the shortcomings of MRI for bedside imaging. However, current 2D ultrasound systems are limited in characterising the soft tissue deformations under mechanical loading because they can only image in the 2D US-plane. As far as the authors are aware of, no studies have quantified how much error results from using a 2D imaging modality (echography) as a substitute for a 3D imaging modality (MRI) for the assessment of soft tissue motion under mechanical loading to inform on injury risk. Moreover, according to figures reported in a National Prevalence Study in French Hospital patients, sacral and heel regions have been reported to be the two most common anatomical sites for PU development (Barrois et al. 2008). Sacrum was therefore chosen in this study as investigation location because of the high prevalence of PUs at this location (Bauer et al. 2016) and ease of access for experimentation.

Because of the significant challenges associated with the *in situ* measurement of both loading and tissue motion using medical imaging in clinical routine, the objective of this of this pilot MRI study was to design an experiment to characterize the sacral soft tissue motion under a controlled mechanical loading. Such an experiment targeted an estimation of the discrepancy between tissue motion assessed using a 2D imaging modality (echography) versus tissue motion assessed using a (reference) 3D imaging modality (Magnetic Resonance Imaging) to inform on the relevance of 2D US-based measurements for pressure ulcer risk assessment. In this work it was not assumed that the out-of-plane movement is negligible.

# Material and Methods

1. Participant

One healthy male volunteer (34 y.o., 1.75m and BMI 27.8 kg/m²) was enrolled in the study after informed consent and local ethics committee agreement (MAP-VS protocol N°ID RCB 2012-A00340-43).

2. MR-compatible custom-made indentation setup

A custom-made MR-compatible setup was designed and assembled with the objective of applying a controlled external compressive force with only a vertical load (i.e. with no shearing load) on the sacrum vertebra via a US-like indenter throughout the MRI acquisitions.

The setup (the reader is referred to Figure 1 for a sketch and to Figure 2 for pictures) was built from a rigid tube structure holding an indenter, a 3D-printed copy of the SL10-2 linear probe transducer developed by (Aixplorer, SuperSonic Imagine, France). The probe was replicated using a Raise3D Pro2 printer and using Raise3D Premium PLA Filament of diameter 1.75 mm (Figure 2 b); final size was approximately 118x28x14 mm$^3$. Indenter connection to the tube structure allowed setting and fixing its orientation. To check if the US plane remained vertical during the experiment, a cylindrical reflective marker (Figure 2 d, Figure 3) was glued on the side of the indenter oriented towards the head of the participant.

Second sacral vertebra (S2) was loaded with the 3D-printed indenter by adding dead weights (plexiglass plates shown in Figure 1 as **J** and **K**) to the structure. The maximum load applied by the setup was determined in accordance to the literature. In the study of (Sheerin and de Frein 2007), the authors reported that pressures measured at the sacral region could reach values as high as 154 mmHg when the head was immobilized and spider strapping was used on an unpadded spinal board. Multiplied by the surface area of the 10-2 linear probe transducer used in this study (410 mm2), this results in a force of approximately 9 N. Preliminary test was performed to evaluate the maximum load the subject could tolerate. Based on this test, a maximum load of 12 N was set for the experiment. This load was applied in 4 loading steps of 200 g or 400 g each resulting in various loads in the range 0-1200 [g] (Table 1).

A schematic representation of the proposed setup with additional details is given in Figure 1.

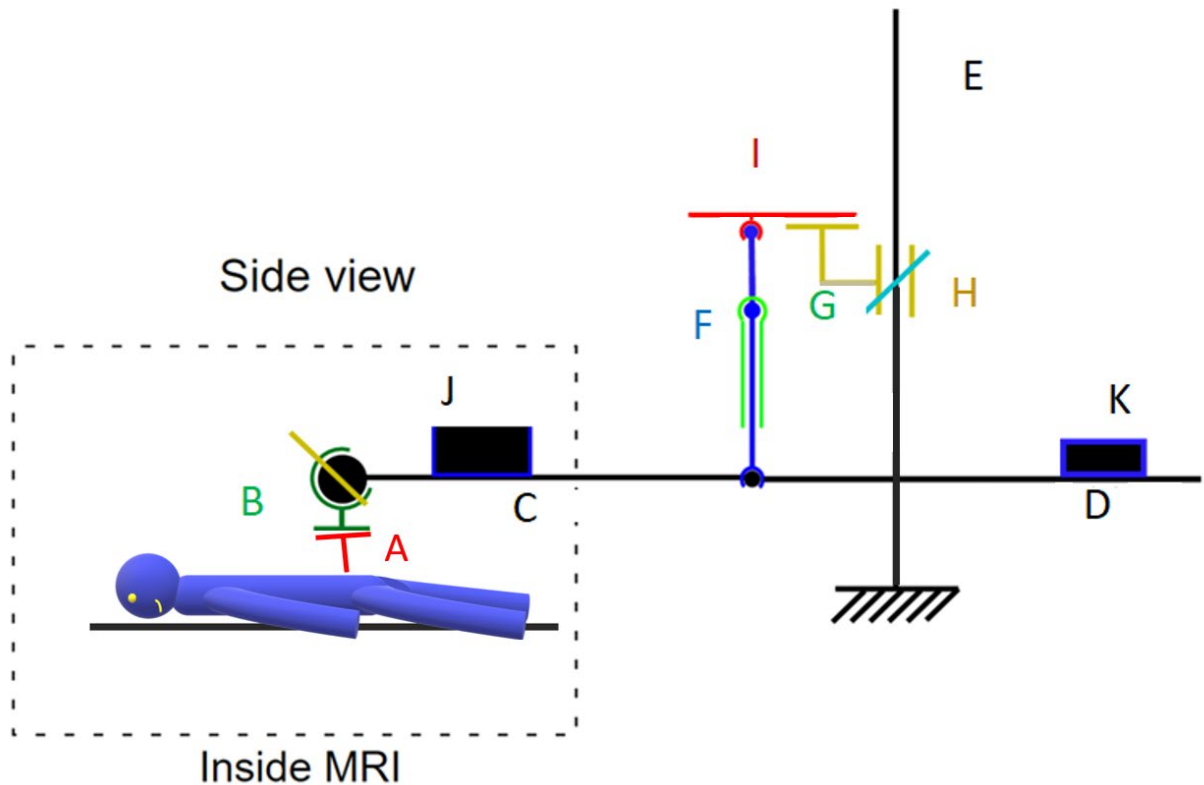

Figure 1 : The contact area **A** indenter/body located at the region of the sacrum with the participant being in a prone position where he was placed head first into the MRI scanner. Clamp **B** holds the indenter, allowing the adjustment of its orientation, to keep it perpendicular to sacrum skin surface. It is positioned close to the left end of the system of rigid tubes (**C** and **D**) which are supported by the MR-compatible stand **E**, also made from rigid tubes. Tube system **C** was built long enough to allow the positioning of the supported indenter inside the MR scanner. To keep the vertical load translation from the setup to the body, the horizontality check of the system of tubes close to the indenter was performed. For the same purpose, the verticality check of two wires **F** holding the tube system **C** was performed using the glass tubes **G** attached to them. The supporting plates **H**, where the wire support **I** was positioned, allow the vertical adjustment on the stand as a way to control the horizontality of the tube system. The loading mass **J**, adjusted for each load case, was positioned close to the indenter to stabilize the load. While the counterweight **K** was positioned on the far end from the indenter to reduce the setup self-weight for the loading step 4.

3. Data collection

3D MRI acquisitions were performed at the IRMaGe research platform (Univ. Grenoble Alpes) in 5 configurations: one unloaded and 4 loaded in decreasing order of loads from 1.2 kg to 400g by steps of 200g or 400g (Figure 2 a, c). A summary of the loads applied is given in Table 1 below.

Prior to the experiment, the uncertainties of the loads applied with the dead weights were assessed by repeating 4 times the previously defined loading scenario and measuring the load with a scale (B3C Sérénite 9260(A), uncertainty of 0.1g). The results are reported as mean +/- 1 SD in Table 1.

The participant was instructed to lie in a prone position in a 3 Tesla Achieva 3.0T dStream Philips scanner (Figure 2 a). No mattress was used during the MRI acquisitions to allow enough space for the loading part of the setup. A soft material was put below the thoracic cage and abdomen so the skin above the first sacral vertebrae was visually horizontal.

For each load case, an MRI volume was acquired. A 3D proton density sequence was used with the following parameters: 399.5x399.5x119.5 $mm^3$ field of view and isotropic voxel size of 0.5x0.5x0.5 $mm^3$. Two surface body coils were placed on either side of the pelvis in the medio-lateral direction (red arrows on Figure 2 a) to increase the signal-to-noise ratio. The acquisition duration time was approximately 10 minutes per MRI scan. Due to limited time allowed in the MRI scanner, approximately 5 minutes were left between acquisitions in order to change the load. Participant was asked to not change body position between the acquisitions. To account for breathing, a gating technique was used.

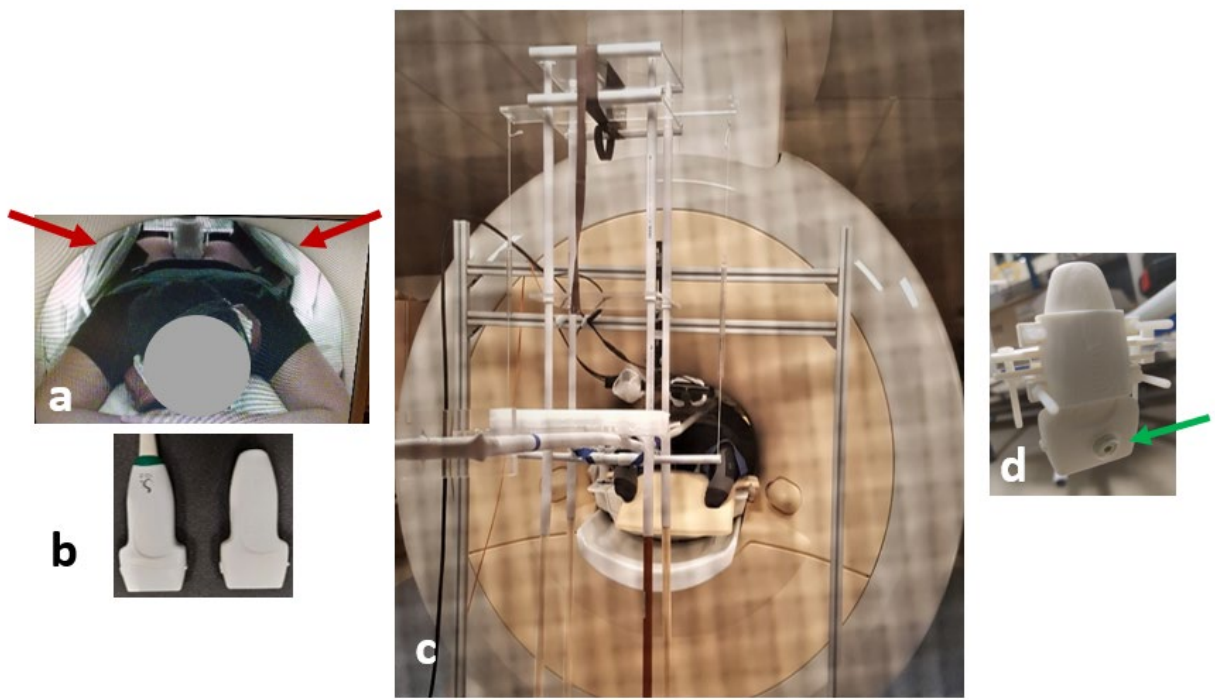

Figure 2 : a) Participant lying in the MR scanner with two surface body coils (red arrows) placed on either side of the pelvis; b) Real US probe (left) and associated 3D-printed copy (right); c) Rear side of the MR scanner showing the rest of the setup and the participant lying in the scanner; d) Green arrow showing the cylindrical reflective marker attached to the 3D-printed indenter.

## 4. Image registration procedure to estimate 3D displacement fields

To extract 3D displacement fields associated with the different loading configurations, a DVC technique was used, based on the open source Elastix library (Klein et al. 2010). The registration was performed between the fixed MRI volume (undeformed configuration, *i.e.* load case 0) and each moving MRI volume (loaded configurations, *i.e.* load cases 1-4).

Voxels of the fixed image were spatially mapped to the voxels of the moving image using a 2-step procedure. First, a rigid body transformation was defined by assuming that the MRI volume is a rigid body. The parameters of the transformation were computed as those that minimized the distance between the bones in the fixed image and each moving image based on a manually defined mask (image segmentation of bone tissues performed with Amira software.) to indicate the bony region assumed non-deformable. Second, B-spline non-rigid transformations were calculated between the aligned moving image and the fixed image. The coefficients of the B-spline transformations were optimized in each cube of a 3D grid that sampled the MR volume (size of the grid: 12mm). Four other grid sizes were tested (8, 10, 14 and 16 mm). However, grid sizes larger than 12 mm resulted in larger differences between the fixed image and the result of registration, while smaller grids introduced noise in some areas of the image. For such optimization, Advanced Normalized Correlation metric was chosen as the similarity measurement since both the fixed and the moving images are obtained using the same modality. Displacements in all directions was calculated with a custom MATLAB script based on the identified displacement field, voxel size and known number of image slices in each direction.

## 5. Data Analysis

Based on the displacement fields obtained by DVC, post processing was performed to estimate the out-of-plane tissue motion. Verticality check of the US image plane was performed.

### 5.1 Construction of the US plane from the reflective marker

The ultrasound image plane (referred to afterwards as "US plane") was constructed as follows: First, two points A and B were manually selected on the posterior side of the reflective marker (Figure 3 b) on a sagittal slice of MR image and used as construction points to define a temporary unit vector $v'_1$. The reflective marker is glued to the indenter surface inclined by 12.1° with respect to the middle transverse plane of indenter (Figure 3 a). A corrected $v_1$ was therefore defined corresponding to a rotation of the vector $v'_1$ so that it lies in the plane parallel to the transverse plane of the indenter. Then, four points C, D, G and H were selected on the superior and inferior sides of the indentation mark on a frontal slice of MR image where the mark was visible as illustrated in Figure 3 below. They were used to define a temporary unit vector $v'_2$ passing through the midpoints of the line segments connecting C and D and G and H respectively (Figure 3 c). Third, a unit vector $v_3$ perpendicular to the plane containing the vectors $v_1$ and $v'_2$ was defined as the result of the cross product ($v_1$ x $v'_2$). A new vector $v_2$ perpendicular to both $v_1$ and $v_3$ was then defined as ($v_3$ x $v_1$).

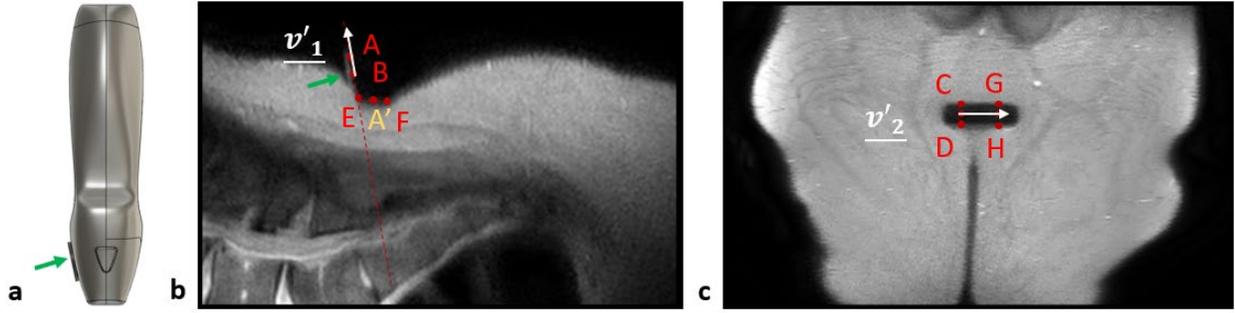

Figure 3 : a) Sagittal view of the 3D printed indenter, green arrow points at the reflective marker; b) Sagittal slice, the red dotted line is aligned with the reflective marker; c) Frontal slice of the MR image.

The US plane was assumed to pass through the middle of the thickness of the piezoelectric transducers. The thickness of the indenter was estimated by selecting 2 points E and F in the same sagittal slice of MR image previously defined (Figure 3 b). The origin of the US coordinate system (US CSYS) was established at the mid-point A' of the line segment joining E and F. Finally, the homogeneous matrix $\underline{\underline{H}}$ of the US CSYS was defined as follows:

$$\underline{\underline{H}} = \begin{bmatrix} v_1 & v_2 & v_3 & A' \\ 0 & 0 & 0 & 1 \end{bmatrix}$$

### 5.2 In-plane and Out-of-plane displacements quantification

To characterize the in-plane and out-of-plane displacements, the displacement field resulting from the 3D Image registration procedure defined in section 4 above was separated into Y-Z (in-plane) and X (out-of-plane) directions in the loading coordinate system (MRI CSYS).

Two parameters were identified to quantify the out-of-plane displacements: First, for each voxel the ratio $R$ of the norm of the out-of-plane $\|D_X\|$ component to the norm of the in-plane component $\|D_{YZ}\|$ was defined. Second, the ratio $N$ of the norm of out-of-plane component $\|D_X\|$ to maximum in-plane component $\max(\|D_{YZ}\|)$ throughout the whole image was defined.

$$R = \frac{abs(D_X)}{\sqrt{D_Y^2 + D_Z^2}} \qquad N = \frac{abs(D_X)}{\max\left(\sqrt{D_Y^2 + D_Z^2}\right)}$$

### 5.3 Regions of interest

Two regions of interest were defined. The first one, called ROI1, was defined in the vicinity of the indentation for the visualization of the displacement field in the soft tissues located above the bony prominence around the indentation zone as depicted in Figure 4 a.

The second, called ROI2, was defined for the evaluation of the out-of-plane displacement according to the definition of ratio $R$ and norm ratio $N$. ROI2 was defined based on the data of load case 1 by selecting the region with the norm ratio values $N$ higher than 0.3. This ROI2 was kept the same for the other load cases as well. An example of ROI1 and ROI2 is given in Figure 4.

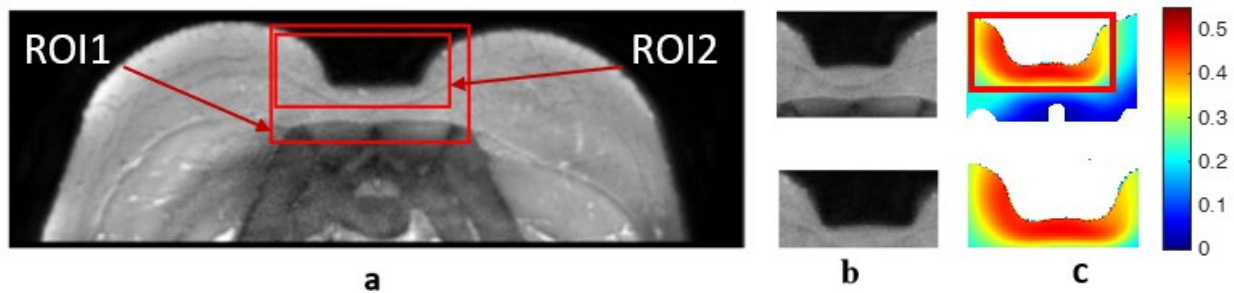

Figure 4 : Two regions of interest. a) Both ROI1 and ROI2 are shown on a transversal MRI slice; b) ROI1 (top) and ROI2 (bottom); c) the norm ratio for the load case 1 plotted for ROI1 (top) and ROI2 (bottom).

### 5.4 Assessment of the uncertainty associated with US plane definition.

To assess the uncertainty associated with the definition of the US CSYS, the points A, B, C, D, E, F, G, H were selected 10 times and associated US CSYS were recalculated. The orientation matrix of the relative angular position of each US CSYS to that of the global MRI CSYS was then calculated. The decomposition of the rotation matrix was done using the YXZ rotation sequences of Cardan angles (MRI CSYS is shown in Figure 6). The rotation of the indenter from the vertical axis was assessed as the angle around the Y axis.

## Results

### 1   Loading uncertainty

The results of the load reproducibility evaluation performed for five load cases are given in Table 1 below.

Table 1 Load case reproducibility evaluation

| Load case | Mean, [g] | Standard deviation (SD), [g] |
|---|---|---|
| L0 | 0 | 0 |
| L1 | 1216 | 2 |
| L2 | 812 | 2 |
| L3 | 626 | 2 |
| L4 | 439 | 2 |

## 2  Data collection results

The MR images of the four loaded configurations (L1 - L4) were analyzed. A transverse slice of MR image in the plane containing the indentation mark left by the 3D-printed indenter of each configuration is given in Figure 5 below.

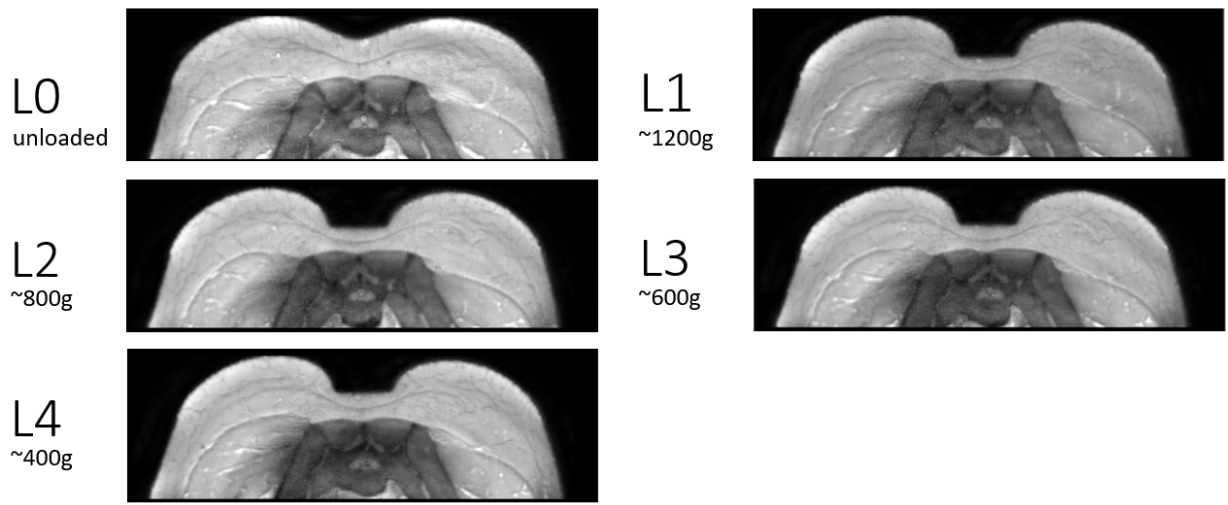

Figure 5 : Transverse slices of the five load cases (L0, L1, L2, L3, L4) at the region of indentation.

## 3. Uncertainty in quantification of US image plane definition.

The mean difference in the relative angular position of each US CSYS to that of the global MRI CSYS was 4.4° (range: [3.0°, 5.8°], standard deviation = 1.2°).

## 4. Displacement fields

### 4.1 In and out-of-plane displacements quantification

The in-plane and the out-of-plane displacements distributions in the ROI1 decomposed in MRI CSYS are shown in Figure 6 for each load case. The same limits were set for all displacement fields. The MRI CSYS was defined with Y and Z being the in-plane directions and X being the out-of-plane direction (Figure 6: first and second rows for in-plane and the last row for out-of-plane correspondingly). A quiver diagram is also drawn to illustrate the combined in-plane displacement distribution in the tissues.

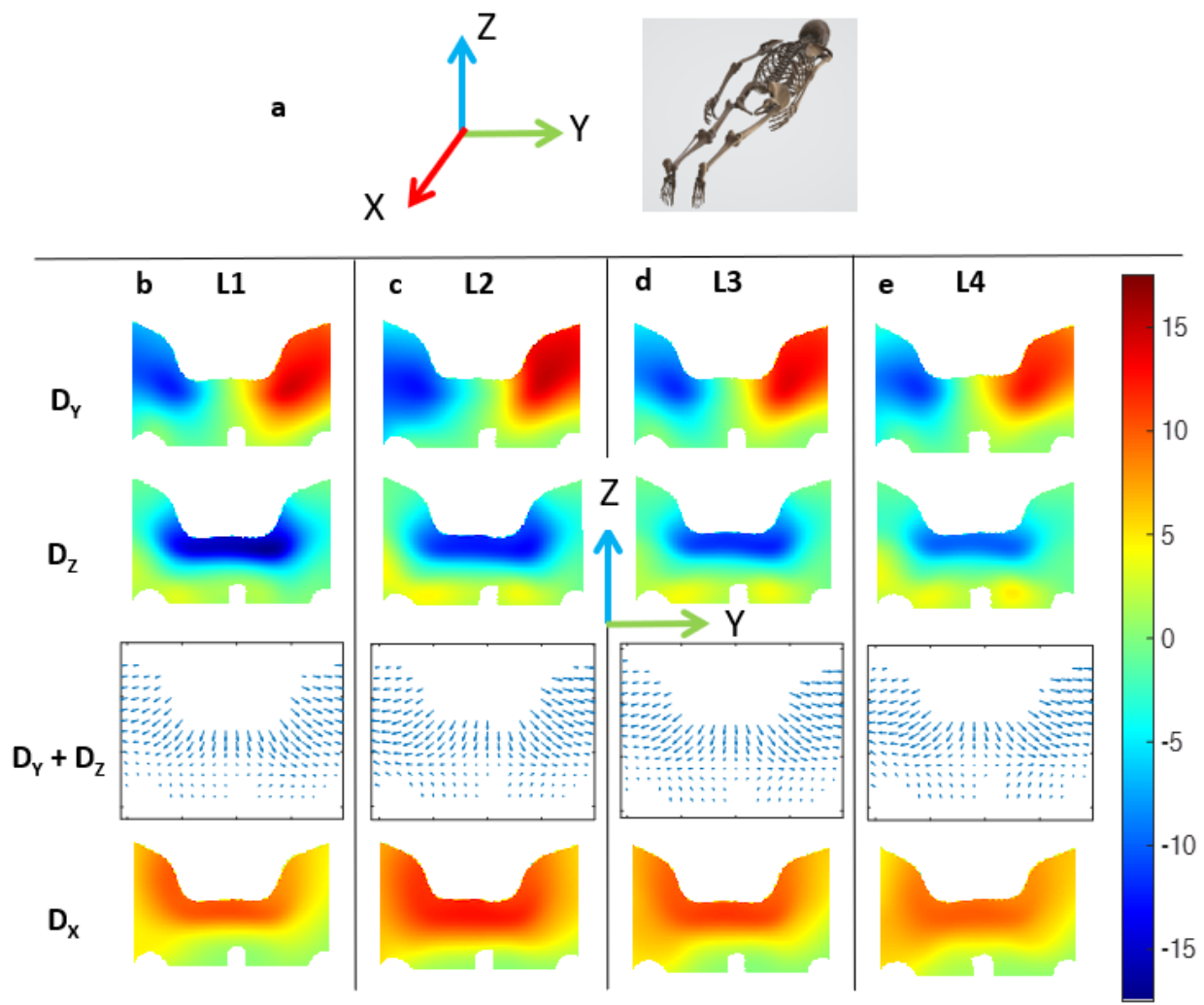

Figure 6 : a) MRI CSYS; b) In-plane Y displacements in the top row; the Z in plane displacements in the second row; the quiver showing the combined in-plane displacements distribution in the third row, and in the last row the out-of-plane X displacements plotted in [mm]; ROI1, L1 (~1200 g) c) L2 (~800 g) d) L3 (~600 g) e) L4 (~400 g).

Table 2 summarized the displacement values for each of the four load cases. When the load was decreased from configuration L1 (~1200 g) to configuration L2 (~800 g), the tissue displacements estimated using DVC in the ROI1 decreased in the Z direction and, on the contrary increased in both the Y and X directions.

When the load was decreased from configuration L2 (800 g) to configuration L3 (600 g), as well as from configuration L3 (600 g) to configuration L4 (400 g), the tissue displacements estimated using DVC decreased in all 3 directions.

| Load case | Max Dx, [mm] | Max Dy, [mm] | Max Dz, [mm] |
|---|---|---|---|
| L1 | 12.4 | 14.6 | 17.5 |
| L2 | 14 | 16.1 | 14.2 |
| L3 | 12.3 | 15.8 | 12.4 |
| L4 | 10.7 | 13.7 | 10.5 |

Table 2 Summary of the maximum displacement values in each of the directions for the four load cases

### 4.2 Parameters of interest to quantify the out-of-plane displacement

The first two rows of Figure 7 present the norm distribution *N* in MRI CSYS on each load case L1 - L4 in the ROI1 and ROI2. The ratio *R* parameter is shown in the ROI2 in the last row. The ratio values increased after the decrease in the load from L1 (~1200g) to L2 (~800g). For all load cases more than half of the voxels had the ratios of the out-of-plane displacement values higher than 0.6, for load cases L2 - L4 the values equal to or higher than 1 were observed in quarter of the voxels of ROI2.

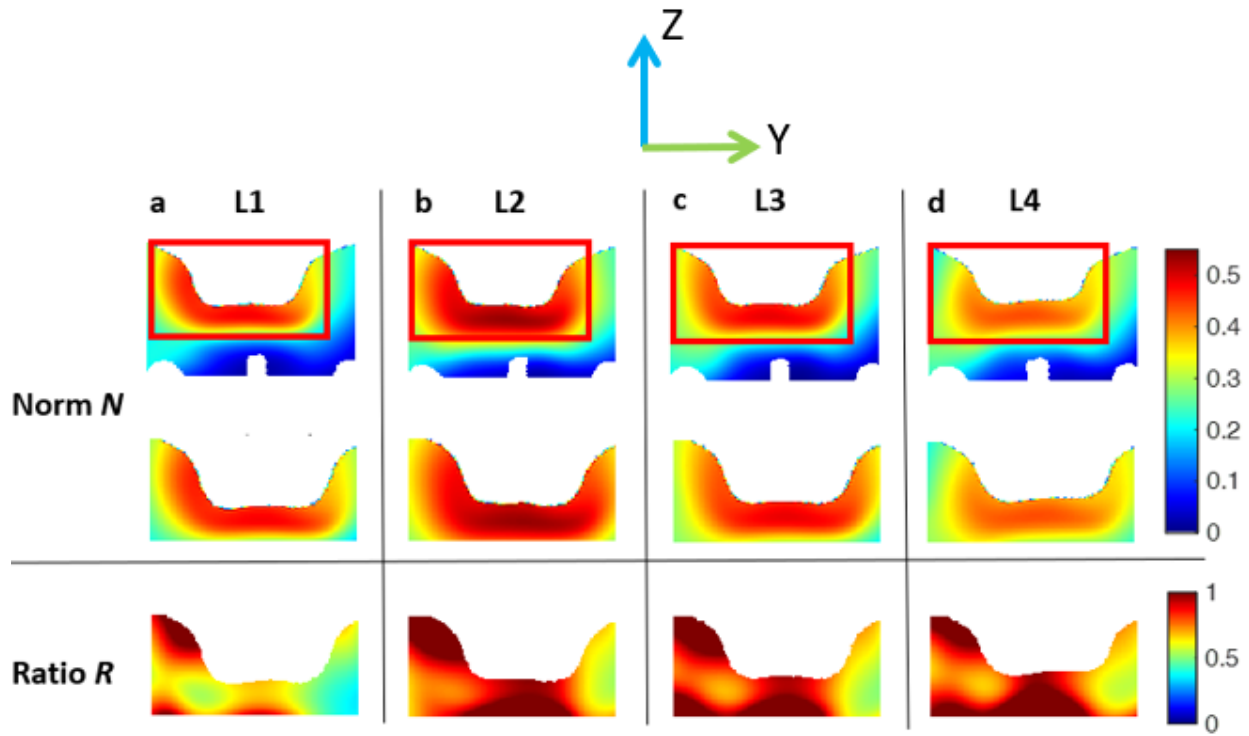

Figure 7 : Norm and ratio of the out-of-plane displacement, ROI2: a) L1 (~1200 g); b) L2 (~800 g); c) L3 (~600 g); d) L4 (~400 g).

## Discussion

The objective of this study was to characterize the out-of-plane displacement using a custom-made MRI-compatible setup under a realistic US transducer loading. This is an important question because the experimental assessment of the internal soft tissue deformation could reveal the individual PU injury risk. However, the assessment with accessible, low cost, and real-time techniques such as 2D B-mode US could be biased by possible out-of-plane motion.

Preliminary results obtained on N=1 healthy volunteer confirm that, for the selected region of interest and chosen loads, the out-of-plane movement is important. The verticality of the US CSYS was also checked: the mean (±SD) value of the angle of interest was 4.4° (±1.2°).

Results showed that the ratios between the out-of-plane and in-plane displacements were higher than 0.6 for more than half of the voxels in the ROI2 for all load cases and higher than 1 for L2-L4 for quarter of the voxels. The inverse correlation between the ratio parameter and the load could be possibly explained by the expulsion of the fluid from the tissues under high deformations. Area of the high norm of the out-of-plane displacement values (more than 0.3 for

L1) is theorized to be limited by the possible shielding effect of the fascia layer located in the adipose tissues.

Variation in the behaviour of the soft tissues in response to the reduction in the load from L1 to L2 was observed in different directions: the displacement in the direction of the loading decreased with the decrease in the load. While in two other directions, the values for L2 were higher than the ones for L1. One possible explanation could be the residual deformation left in the tissues.

There are some limitations to the current study. First, only one healthy male volunteer was recruited for this study because of the difficulty of access to MRI. The sample is not really representative of neither at-risk sub-population ((Coleman et al. 2013) showed that most endangered are elderly, bedridden or SCI subjects) nor the anatomic gender differences (especially in the pelvic region where the tissue organization varies). Results of a secondary data analysis performed by Kottner et al. suggest that underweight patients (with BMI < 18.5 kg/m$^2$) are more susceptible to trunk PU development than patients with normal weight or overweight (Kottner, Gefen, and Lahmann 2011). While the BMI of the participant of the current study was 27.4 kg/m$^2$, the conclusions of this preliminary study should therefore be considered specific to this application and further work aiming at generalizing these results needs to be performed. Second limitation is related to the artificially designed loading conditions. One of the main prerequisites was to have a control over the value and the direction of loading; therefore, it was decided to limit the experiment only to vertical compressive load. While in the clinical environment, region of the sacrum is highly affected by shear loads. The possibility of the controlled shear loading will be investigated in future studies. Choice of the order of the load steps (from the high load to the lower one) as well as a choice of the registration parameters (metric, grid size) for the non-rigid transformations might have had an effect on the results. An Advanced Normalised Correlation similarity metric was used for the DVC while the other available similarity measurements were not tested, this constitutes a perspective work. Another bottleneck is the validation of the registration method; in current case only a visual assessment was performed.

DVC-MRI combination was previously used in 3D modality for the *ex vivo* assessment of tissue motion in the intervertebral discs (Tavana et al. 2020), in rodent lungs (Arora et al. 2021) and *in vivo* in human lower leg muscles (Yaman et al. 2013) and in the heel (Trebbi et al. 2021). Another 3D combination, DVC-US, combining the accessibility of the US with the completeness of the tissue motion data in 3D, was also investigated in the literature. *In vivo* algorithms of DVC-US were tested in human liver (Foroughi, Abolmaesumi, and Hashtrudi-Zaad 2006) and heart

(Shekhar et al. 2004). This combination of DVC and 3D US is a promising future research topic in application to sacral region.

## Conclusion

Several limitations have been identified as a perspective work such as the lack of clinically relevant loading and inclusion of only one healthy subject. However, if confirmed in larger sample, the implications of our results could go far beyond the scope of the PU prevention. While not providing the information about the out-of-plane tissue movement and anisotropy, US imaging and associated correlation techniques are indeed used to investigate biomechanical properties of various soft tissues including fascia (Langevin et al. 2011), the intervertebral disc (Vergari et al. 2014) and thigh (Fougeron et al. 2020). Results of the current study suggest that the 2D US should be used with caution for the evaluation of tissue motion. Possibility of using 3D US as an accessible alternative to 3D MRI modality should be further investigated in application to sacrum region.

## Acknowledgements

This project has received funding from the European Union's Horizon 2020 research and innovation programme under the Marie Skłodowska-Curie grant agreement No. 811965. IRMaGe MRI facility was partly funded by the French program "Investissement d'Avenir" run by the "Agence Nationale pour la Recherche"; grant "Infrastructure d'avenir en Biologie Sante" - ANR-11-INSB-0006.

# References

Affagard, J. S., P. Feissel, and S. F. Bensamoun. 2015. "Measurement of the Quadriceps Muscle Displacement and Strain Fields with Ultrasound and Digital Image Correlation (DIC) Techniques." *Irbm* 36 (3): 170–77. https://doi.org/10.1016/j.irbm.2015.02.002.

Arora, Hari, Ria L. Mitchell, Richard Johnston, Marinos Manolesos, David Howells, Joseph M. Sherwood, Andrew J. Bodey, and Kaz Wanelik. 2021. "Correlating Local Volumetric Tissue Strains with Global Lung Mechanics Measurements." *Materials* 14 (2): 1–17. https://doi.org/10.3390/ma14020439.

Barrois, B., C. Labalette, P. Rousseau, A. Corbin, D. Colin, F. Allaert, and J. L. Saumet. 2008. "A National Prevalence Study of Pressure Ulcers in French Hospital Inpatients." *Journal of Wound Care* 17 (9). https://doi.org/10.12968/jowc.2008.17.9.30934.

Bauer, Karen, Kathryn Rock, Munier Nazzal, Olivia Jones, and Weikai Qu. 2016. "Pressure Ulcers in the United States' Inpatient Population From 2008 to 2012: Results of a Retrospective Nationwide Study." *Ostomy Wound Management* 62 (11): 30–38.

Bay, B. K. 2008. "Methods and Applications of Digital Volume Correlation." *Journal of Strain Analysis for Engineering Design* 43 (8): 745–60. https://doi.org/10.1243/03093247JSA436.

Ceelen, K. K., A. Stekelenburg, J. L.J. Mulders, G. J. Strijkers, F. P.T. Baaijens, K. Nicolay, and C. W.J. Oomens. 2008. "Validation of a Numerical Model of Skeletal Muscle Compression with MR Tagging: A Contribution to Pressure Ulcer Research." *Journal of Biomechanical Engineering* 130 (6): 1–8. https://doi.org/10.1115/1.2987877.

Chernak Slane, Laura, and Darryl G. Thelen. 2014. "The Use of 2D Ultrasound Elastography for Measuring Tendon Motion and Strain." *Journal of Biomechanics* 47 (3): 750–54. https://doi.org/10.1016/j.jbiomech.2013.11.023.

Chimenti, Ruth L., A. Samuel Flemfister, John Ketz, Mary Bucklin, Mark R. Buckley, and Michael S. Richards. 2016. "Ultrasound Strain Mapping of Achilles Tendon Compressive Strain Patterns During Dorsiflexion." *Journal of Biomechanics* 49 (1): 39–44. https://doi.org/10.1016/j.jbiomech.2015.11.008.Ultrasound.

Coleman, Susanne, Claudia Gorecki, E. Andrea Nelson, S. José Closs, Tom Defloor, Ruud Halfens, Amanda Farrin, Julia Brown, Lisette Schoonhoven, and Jane Nixon. 2013. "Patient Risk Factors for Pressure Ulcer Development: Systematic Review." *International Journal of Nursing Studies* 50 (7): 974–1003. https://doi.org/10.1016/j.ijnurstu.2012.11.019.

Doridam, J., A. Macron, C. Vergari, A. Verney, P. Y. Rohan, and H. Pillet. 2018. "Feasibility of Sub-Dermal Soft Tissue Deformation Assessment Using B-Mode Ultrasound for Pressure Ulcer Prevention." *Journal of Tissue Viability* 27 (4): 238–43. https://doi.org/10.1016/j.jtv.2018.08.002.

Foroughi, Pezhman, Purang Abolmaesumi, and Keyvan Hashtrudi-Zaad. 2006. "Intra-Subject Elastic Registration of 3D Ultrasound Images." *Medical Image Analysis* 10 (5): 713–25. https://doi.org/10.1016/j.media.2006.06.008.

Fougeron, Nolwenn, Pierre-Yves Rohan, Diane Haering, Jean-Loïc Rose, Xavier Bonnet, and Hélène Pillet. 2020. "Combining Freehand Ultrasound-Based Indentation and Inverse Finite Element Modelling for the Identification of Hyperelastic Material Properties of Thigh Soft Tissues." *Journal of Biomechanical Engineering*, no. c: 1–22. https://doi.org/10.1115/1.4046444.

Gawlitta, Debby, Wei Li, Cees W.J. Oomens, Frank P.T. Baaijens, Dan L. Bader, and Carlijn V.C. Bouten. 2007. "The Relative Contributions of Compression and Hypoxia to Development of Muscle Tissue Damage: An in Vitro Study." *Annals of Biomedical Engineering* 35 (2): 273–84. https://doi.org/10.1007/s10439-006-9222-5.

Gennisson, J. L., T. Deffieux, M. Fink, and M. Tanter. 2013. "Ultrasound Elastography: Principles and Techniques." *Diagnostic and Interventional Imaging* 94 (5): 487–95. https://doi.org/10.1016/j.diii.2013.01.022.

Gilchrist, Christopher L., Jessie Q. Xia, Lori A. Setton, and Edward W. Hsu. 2004. "High-Resolution Determination of Soft Tissue Deformations Using MRI and First-Order Texture Correlation." *IEEE Transactions on Medical Imaging* 23 (5): 546–53. https://doi.org/10.1109/TMI.2004.825616.

Klein, Stefan, Marius Staring, Keelin Murphy, Max A. Viergever, and Josien P.W. Pluim. 2010. "Elastix: A Toolbox for Intensity-Based Medical Image Registration." *IEEE Transactions on Medical Imaging* 29 (1): 196–205. https://doi.org/10.1109/TMI.2009.2035616.

Kottner, Jan, Amit Gefen, and Nils Lahmann. 2011. "Weight and Pressure Ulcer Occurrence: A Secondary Data Analysis." *International Journal of Nursing Studies* 48 (11): 1339–48. https://doi.org/10.1016/j.ijnurstu.2011.04.011.

Langevin, Helene M., James R. Fox, Cathryn Koptiuch, Gary J. Badger, Ann C. Greenan- Naumann, Nicole A. Bouffard, Elisa E. Konofagou, Wei Ning Lee, John J. Triano, and Sharon M. Henry. 2011. "Reduced Thoracolumbar Fascia Shear Strain in Human Chronic Low Back Pain." *BMC Musculoskeletal Disorders*. https://doi.org/10.1186/1471-2474-12-203.

Loerakker, S., E. Manders, G. J. Strijkers, K. Nicolay, F. P.T. Baaijens, D. L. Bader, and C. W.J. Oomens. 2011. "The Effects of Deformation, Ischemia, and Reperfusion on the Development of Muscle Damage during Prolonged Loading." *Journal of Applied Physiology* 111 (4): 1168–77. https://doi.org/10.1152/japplphysiol.00389.2011.

Loerakker, S. S., A. Stekelenburg, G. J. Strijkers, J. J.M. Rijpkema, F. P.T. Baaijens, D. L. Bader, K. Nicolay, and C. W.J. Oomens. 2010. "Temporal Effects of Mechanical Loading on Deformation-Induced Damage in Skeletal Muscle Tissue." *Annals of Biomedical Engineering* 38 (8): 2577–87. https://doi.org/10.1007/s10439-010-0002-x.

Nelissen, Jules L., Ralph Sinkus, Klaas Nicolay, Aart J. Nederveen, Cees W.J. Oomens, and Gustav J. Strijkers. 2019. "Magnetic Resonance Elastography of Skeletal Muscle Deep Tissue Injury." *NMR in Biomedicine* 32 (6): 1–12. https://doi.org/10.1002/nbm.4087.

Nierop, Bastiaan J. van, Anke Stekelenburg, Sandra Loerakker, Cees W. Oomens, Dan Bader, Gustav J. Strijkers, and Klaas Nicolay. 2010. "Diffusion of Water in Skeletal Muscle Tissue Is Not Influenced by Compression in a Rat Model of Deep Tissue Injury." *Journal of Biomechanics* 43 (3): 570–75.

https://doi.org/10.1016/j.jbiomech.2009.07.043.

Sheerin, Fintan, and Ruairi de Frein. 2007. "The Occipital and Sacral Pressures Experienced by Healthy Volunteers Under Spinal Immobilization: A Trial of Three Surfaces." *Journal of Emergency Nursing* 33 (5): 447–50. https://doi.org/10.1016/j.jen.2006.11.004.

Shekhar, Raj, Vladimir Zagrodsky, Mario J. Garcia, and James D. Thomas. 2004. "Registration of Real-Time 3-D Ultrasound Images of the Heart for Novel 3-D Stress Echocardiography." *IEEE Transactions on Medical Imaging* 23 (9): 1141–49. https://doi.org/10.1109/TMI.2004.830527.

Solis, Leandro R., Adrian B. Liggins, Peter Seres, Richard R.E. Uwiera, Niek R. Poppe, Enid Pehowich, Richard B. Thompson, and Vivian K. Mushahwar. 2012. "Distribution of Internal Strains around Bony Prominences in Pigs." *Annals of Biomedical Engineering* 40 (8): 1721–39. https://doi.org/10.1007/s10439-012-0539-y.

Sonenblum, Sharon Eve, Stephen H. Sprigle, John Mc Kay Cathcart, and Robert John Winder. 2015. "3D Anatomy and Deformation of the Seated Buttocks." *Journal of Tissue Viability* 24 (2): 51–61. https://doi.org/10.1016/j.jtv.2015.03.003.

Stekelenburg, A., C. W.J. Oomens, G. J. Strijkers, K. Nicolay, and D. L. Bader. 2006. "Compression-Induced Deep Tissue Injury Examined with Magnetic Resonance Imaging and Histology." *Journal of Applied Physiology* 100 (6): 1946–54. https://doi.org/10.1152/japplphysiol.00889.2005.

Tavana, S., J. N. Clark, J. Prior, N. Baxan, S. D. Masouros, N. Newell, and U. Hansen. 2020. "Quantifying Deformations and Strains in Human Intervertebral Discs Using Digital Volume Correlation Combined with MRI (DVC-MRI)." *Journal of Biomechanics* 102: 109604. https://doi.org/10.1016/j.jbiomech.2020.109604.

Traa, Willeke A., Mark C. van Turnhout, Kevin M. Moerman, Jules L. Nelissen, Aart J. Nederveen, Gustav J. Strijkers, Dan L. Bader, and Cees W.J. Oomens. 2018. "MRI Based 3D Finite Element Modelling to Investigate Deep Tissue Injury." *Computer Methods in Biomechanics and Biomedical Engineering* 21 (14): 760–69. https://doi.org/10.1080/10255842.2018.1517868.

Traa, Willeke A., Mark C. van Turnhout, Jules L. Nelissen, Gustav J. Strijkers, Dan L. Bader, and Cees W.J. Oomens. 2019. "There Is an Individual Tolerance to Mechanical Loading in Compression Induced Deep Tissue Injury." *Clinical Biomechanics* 63: 153–60. https://doi.org/10.1016/j.clinbiomech.2019.02.015.

Trebbi, Alessio, Antoine Perrier, Mathieu Bailet, and Yohan Payan. 2021. "MR-Compatible Loading Device for Assessment of Heel Pad Internal Tissue Displacements under Shearing Load." *Medical Engineering and Physics* 98 (November): 125–32. https://doi.org/10.1016/j.medengphy.2021.11.006.

Vergari, Claudio, Philippe Rouch, Guillaume Dubois, Dominique Bonneau, Jean Dubousset, Mickael Tanter, Jean Luc Gennisson, and Wafa Skalli. 2014. "Intervertebral Disc Characterization by Shear Wave Elastography: An in Vitro Preliminary Study." *Proceedings of the Institution of Mechanical Engineers, Part H: Journal of Engineering in Medicine* 228 (6): 607–15. https://doi.org/10.1177/0954411914540279.

Yaman, Alper, Cengizhan Ozturk, Peter A. Huijing, and Can A. Yucesoy. 2013. "Magnetic Resonance Imaging Assessment of Mechanical Interactions between Human Lower Leg Muscles in Vivo." *Journal of Biomechanical Engineering* 135 (9). https://doi.org/10.1115/1.4024573.

Zhu, Haijiang, Shifeng Zhou, Ping Yang, Longbiao He, and Jinglin Zhou. 2015. "An Efficient Optimal Method for a 2D Strain Estimation of Ultrasound Tissue-Mimicking Material Phantom." *Polymer Engineering and Science* 55 (12): 2751–60. https://doi.org/10.1002/pen.